\documentclass{aa} 
\usepackage{graphicx}
\graphicspath{{figures/}}
\usepackage[varg]{txfonts}
\usepackage{bm}         
\usepackage{rotating}
\usepackage{color}
\usepackage{xcolor}
\usepackage{color} 
\usepackage{footnote}
\makesavenoteenv{tabular}
\usepackage[toc,page]{appendix}
\usepackage{hyperref} 
\usepackage{ulem}

\newcommand{\Msun}{M$_\odot$}

\newcommand{\nifs}{$^{56}$Ni}
\newcommand{\Nimass}{{M($^{56}\mathrm{Ni}$)}}

\definecolor{yaleblue}{rgb}{0.1,0.3,0.9}
\definecolor{lava}{rgb}{0.81, 0.06, 0.13}
\definecolor{forestgreen}{rgb}{0.0, 0.27, 0.13}
\hypersetup{colorlinks=true, linkcolor=lava, urlcolor=forestgreen, citecolor=yaleblue}
\begin{document} 

\title{Supernovae double-peaked light curves from double-nickel distribution}
\author{Mariana Orellana\inst{1,2}, 
Melina C. Bersten\inst{3,4,5}}
\institute{Universidad Nacional de R\'{\i}o Negro. Sede Andina, Mitre 630 (8400) Bariloche, Argentina
\and  Consejo Nacional de Investigaciones Científicas y Técnicas (CONICET), Argentina
\and  Instituto de Astrof\'\i sica de La Plata (IALP), CCT-CONICET-UNLP. Paseo del Bosque S/N (B1900FWA), La Plata, Argentina
\and  Kavli Institute for the Physics and Mathematics of   the Universe (WPI), The University of  Tokyo, 5-1-5 Kashiwanoha, Kashiwa, Chiba 277-8583, Japan
\and  Facultad de Ciencias Astron\'omicas y Geof\'{\i}sicas, Universidad Nacional de La Plata, Paseo del Bosque S/N, B1900FWA La Plata, Argentina}
\offprints{M. Orellana\\ \email{morellana@unrn.edu.ar}}

\titlerunning{double SNe -- double Ni}

\authorrunning{Orellana \& Bersten}
\date{Received 26 May 2022; accepted 26 August 2022}
\abstract {Among supernovae (SNe) of different luminosities, many double-peaked light curves (LCs) have been observed, representing a broad morphological variety. In this work, we investigate which of these can be modelled by assuming a double-peaked distribution of their radioactive material, as originally proposed for SN2005bf. The inner zone corresponds to the regular explosive nucleosynthesis and extends outwards, according to the usual scenario of mixing. The outer \nifs--rich shell may be related to the effect of jet-like outflows that have interacted  with more distant portions of the star before the arrival of the SN shock. As the outer layer is covered by matter that is optically less thick, its energy emerges earlier and generates a first peak of radiation.
To investigate this scenario in more detail, we have applied our hydrodynamic code that follows the shock propagation through the progenitor star and takes into account the effect of the $\gamma$-ray photons produced by the decay of the radioactive isotopes. 
We present a simple parametric model for the \nifs\ abundance profile and explore the consequences on the LC of individually varying the quantities that define this distribution, setting our focus onto the stripped-envelope progenitors. In this first study, we are interested in the applicability of this model to SNe that have not been classified as superluminous, thus, we have selected our parameter space accordingly.
Then, within the same mathematical prescription for the \nifs -profile, we revisited the modelling process for a series of objects: SN2005bf, PTF2011mnb, SN2019cad, and SN2008D. In some cases, a decrease in the gamma ray opacity is required to fit the late time observations. We also discuss the other cases in which this scenario might be likely to explain the LC morphology. A steep initial decline in the observed bolometric LC within less than few days after the explosion becomes less feasible for this model, because it requires a large abundance of \nifs\ near the stellar surface, indicating a strongly inverted distribution.
An initial bolometric rise before the two peaks seems more favourable for the double-nickel case, particularly as it can be difficult to explain through other scenarios, unless a combination of power sources is invoked.}

\maketitle
\section{Introduction} 
\label{sec:intro}
The observed shape of the LC of H-poor core collapse SNe (or stripped envelope supernovae) usually exhibits only one maximum in the bolometric luminosity. 
This maximum is well understood as being powered by the decay of radioactive material generated during the explosion

The ability to detect SNe at increasingly early times has been evolving a great deal in recent years (see e.g. projects such as ZTF, KISS, and HITS, among others). Thus, there is also a higher fraction of SNe with early detection. 
From theoretical grounds an early cooling phase is predicted, consisting of a decrease in bolometric luminosity prior to the main peak. 
The duration of this phase depends on the progenitor structure, for a more compact star, the faster the associated cooling phase 
\citep[see e.g.][]{2012Bersten}.
Therefore, the detectability of the cooling phase depends on the SNe type. For example, for type IIb SNe that are expected to retain some relatively extended H-rich envelope the cooling phase has been more easily observed as opposed to SNe Ib or Ic for whom this phase has been elusive. In very few cases, some emission for SNe Ib or Ic has been reported and generally the detection is only in one or two photometric bands (see e.g. the case of SN 1999ex, \citealt{2002Stritzinger}, or iPTF15dtg, \citealt{2016Taddia}).
In addition to the cooling emission there are some cases where two clear peak has been discovered. Currently, there is a growing variety of double-peaked supernovae that have been reported, however, they are still uncommon.

The peculiar types of SNe, along with the basic premises of how they can be distinguished, are presented in \cite{2017Kasen}. Naturally, more rare events have been discovered since the advent and growth of wide-field surveys and rapid-response follow-up facilities \citep[see][and references therein]{2019Modjaz}. 

Throughout this work, we will refer to the peaks of luminosity that occur apart from the spike of the shock breakout (SBO). The SBO is expected to happen even before the cooling emission and is harder to detect
(see e.g. the SBO detection of SN 2016gkg, \citealt{2018Bersten}).

The luminosities of double-peaked light curves (LCs) have shown a wide range, including superluminous SNe.
The frequency and detectability of double-peaked SLSNe was examined by \cite{2016Nicholl} based on the flux excess from polynomial fits at the earliest epochs. After an intensive literature search, we do not know of a similar study for double-peaked SNe with normal luminosities.

The ideas to explain the double maxima of the LC mainly comprise the interaction between the ejecta and either a modified structure at the outskirt of the progenitor or with unbounded circumstellar matter (CSM) surrounding the progenitor \citep{2013Moriya, 2015Piro,2020Ertl, 2021Jin} and with different extents \citep{2012Bersten, Nakar2014,2016Taddia}. 
This interpretation works fairly well when the observed brightness starts by decreasing (cooling phase) before rising to the radioactive peak. 
However, that scenario faces a challenge in cases where the luminosity is initially increasing to a first maximum that takes place later than about 10 days since the explosion.

For certain LCs, a promising alternative involve nickel plus a tunable central engine, which is likely a magnetar that produces a delayed injection of energy or models in which the progenitor star expelled shells of material years before the explosion and get detached from the star, so the interaction produces an important rebrightening at late phases (see e.g. \citealt{2020Li}).
Another possibility that was proposed and calculated a decade ago concerns an early feature that may result from the ejecta colliding with a binary companion, granted it is a system with a suitable inclination \citep{2010Kasen}.

The unprecedented LC morphology among Type Ib/Ic of SN2005bf \citep{2005Tominaga}, with  a first maximum about 16 days after explosion and a second brighter peak 24 days later, drives the proposal of an unusual \nifs\ distribution in the model developed by \cite{Folatelli06}. 
That concept inspired an alternative explanation for SN2008D with the scenario consisting of \nifs -rich material deposited in the outer layers of the ejecta, carried out by a jet-like phenomenon \citep{2013Bersten}, which is quite consistent with the X-ray emission early detected in association with this SN \citep{2008Soderberg,Mazzali2008}. 
Later, LSQ13abf was found to resemble SN2008D and was subject to a similar analysis with regard to the justification of the shape of the LC \citep{2020Stritzinger}.
PFT11mnb and SN2019cad showed stronger analogies with the LC of SN2005bf and, therefore, the model with a double \nifs\ profile was also proposed for them
\citep{Taddia18,Gutierrez2021}.
We also refer to \cite{2020Magee}, who recently explored the case of \nifs\ layers as a source of early light curve bumps in type Ia supernovae.
The external nickel that forms a high-velocity blob is the source of extra energy needed to explain the first peak. This model has the advantage of having a single power source, but at the cost of considering a more elaborate distribution. However, aside from a few references, its potential has not been deeply explored and this is the motivation for the present study. 

This paper is organized as follows: first, we present the numerical setup of the 1D radiation-hydrodynamic simulations in Section~\ref{sect:models} and apply it to a fixed progenitor.
The resulting parameter exploration is included in Section \ref{space}.
We then present the information recollected from observed SNe with early detections in Section~\ref{sec:observadas}. 
In Section \ref{family}, we present the modelling of our selected family, along with a brief comparison to previous results.
A more extreme example is presented in Section~\ref{08D}, demonstrating an earlier first peak. 
We conclude our paper with a discussion in Section~\ref{sec:discuss}.


\section{Numerical models}
\label{sect:models}

We used a one-dimensional (1D) local thermal equilibrium (LTE) hydrodynamical code to perform our numerical exploration. The code assumes the diffusion approximation for optical photons. 
For the gamma rays produced through the radioactive decay, we applied a gray approximation. 
A constant value of $\kappa_\gamma =0.03$ cm$^2$/g was adopted for the gamma opacity. That quantity was modified in some cases in order to model the last decline in the LC (discussed in Section~\ref{family}). All the details on the 
 equations of state and opacities tables adopted in the code can be found in \cite{Bersten2011}. The explosion is simulated by injecting energy near to the progenitor core which produces a powerful shock wave that propagates inside the ejecta until its arrives to the surface where the photons can begin to propagate out, producing the first electromagnetic signal of the SN. Therefore, the code computes the whole evolution of the LC in a self consistent way up to the nebular phase. 
However, due to the crude treatment of the radiation transfer, even if the broadband photometry can be estimated by assuming a blackbody emission, only the bolometric luminosity is confidently calculated. Thus, our analysis is focussed on the properties of bolometric LCs.

An initial configuration in hydrostatic equilibrium was used to initiate the hydrodynamical calculations. In this study, we employed progenitor models from stellar evolutionary calculations (more details given below), however, the distribution of the radioactive material, which is in principle a consequence of the explosive burning, is treated in a parametric way and assumed to be an initial condition in our calculation. 
This has been extensively applied in the literature where the nickel is usually mixed out by hand, justified by the effects of instabilities which cannot be properly taken into account in 1D calculations. 
Since our code allows for any distribution of this material inside the ejecta, here, we go one step forward and we assume that the nickel can have two separate rich regions: an inner component which is typically assumed in SN models and an external component. Here, we are interested in exploring the LC morphology produced with this type of double-nickel profile in a parametric
and systematic way.

Although a self-consistent model is beyond the scope of this study, the double-peaked \nifs\ distribution is well motivated. It has proven to be successful in modelling a few SNe and suggested for some others (see Section \ref{sec:observadas})
It is interesting to have them re-investigated within a matching parameterization for the \nifs -profile.
The framework is established by the hypothetical presence of outflows involved in the explosion, as in the case of gamma-ray burst (see e.g. \citealt{MacFadyen2001, 2006Woosley, Banerjee13}). One possible effect of the jet-like outflows is to trigger the nucleosynthesis of radioactive elements somewhere at the outer layers of the ejecta before the shock front of the SN arrives \citep{2015Nishimura}. Alternatively, perhaps the jets can transport out some material mixed with radioactive elements. We speculate that such a displacement could be more efficient than hydrodynamical instabilities usually invoked for the mixing, which justifies the hypothesis of a separate nickel-rich layer.
The presence of jet-like outflows does not imply an energetic photon detection.
The high-energy emission might not occur because the jets do not break the stellar surface, although, alternatively, even if they do so, the energetic photons may not be detected for geometric reasons \citep{2018bMargalita}.
In fact, a GRB is thought to be produced in the unusual case, where the jet tunnels throughout the entire star and is viewed on-axis \citep{2017Sobacchi}.

Details of the jet dynamics, the interaction with the stellar matter, and nucleosynthesis are not specified in our model. Throughout this work, we assume that the external \Nimass\ component is a fraction of the one sited at the inner layers.
Also, we restrain the amount of \nifs\ within the range of previously published mass values in normal stripped-envelope SNe, namely, a mean value of 0.293~\Msun\ and ranging between $0.03 - 2.4$~\Msun\ \citep{Anderson2019}. We consider the upper part of this range to be an extreme overestimate for the normal SNe, as it is more than $3\sigma$ aside from the mean value. We also note that this author has remarked on the questionable accuracy of popular methods used to estimate \Nimass.

A double-peaked nickel distribution was assumed for the first time to model the unusual LC morphology of SN 2005bf. 
The two components presented in that study follows a continuous profile with abundance values always greater than zero and denoted as a mildly inverted distribution (see details in \cite{Folatelli06}). Here, we preferred to assume a more simplistic profile with two boxes for the nickel abundance, with a mass fraction denoted by $X$. In our numerical scheme,  $X$ is a function of Lagrangian mass coordinate ($M_r$), which is directly mapped to the mass fraction, $f$, a more general variable that allows for comparisons between different progenitors.  Although we decided to use a simplified nickel profile, we explored more complex profiles to facilitate the interpretation of the effect of each parameter on the LC. In particular, we tested the effect of assuming smoother \nifs\ profiles, but we have not found any noticeable difference in the resulting LC, as we discuss in \cite{Gutierrez2021}. 

In the next section (Section~\ref{space}), we present and discuss our results for a fixed progenitor star with a pre-SN mass of 11 \Msun\ (denoted {\tt He11}) and a radius of $\sim 5$~R$_\odot$. This structure corresponds to a main sequence star of 30 \Msun\ and its evolution was computed with the public stellar evolution \textsc{MESA} code \citep{2011Paxton} with no rotation and other setup as described in \cite{Martinez2020}.

\section{Simple double \nifs\ parameter space}\label{space}

\begin{figure}
\resizebox{\hsize}{!}{\includegraphics{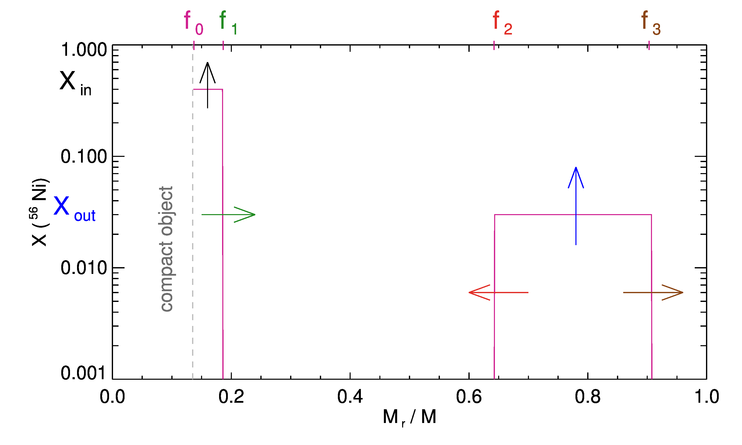}}
\caption{
Schematic abundance profile of \nifs\ as a function of the interior mass fraction in the star. The parameters used to describe this distribution are indicated in the axes. The color of each arrow and corresponding label is used in next figures to facilitate the interpretation.} 
\label{fig:params}
\end{figure}

Figure~\ref{fig:params} displays the scheme of the nickel distribution analyzed here. The profile has an inner \nifs-rich layer and an outer component, both with constant nickel abundances denoted by $X_{\rm in}$ and $X_{\rm out}$, respectively. 
These layers are located at separate zones, with no nickel in the middle, and the spread is regulated by fractions, $f,$ of the total mass, $M,$ of the progenitor. 
The abundances of all the other elements in a given shell add $(1-X)$ to fulfill the normalization.
Once the SN reaches the homologous expansion, this distribution in mass could be expressed as a function of velocity, as done in \cite{2013Bersten}, with the notion of a slower (for the inner) and a faster  (for the external) \nifs -rich shells.
We considered an inner component close to the compact remnant, namely, extending from the inner border of the ejecta, $f_0$, up to a fraction, $f_1$, that has the usual partial-mixing sense when the nickel abundance profile is a boxcar function (as in \citealt{Bersten2011}, though different from e.g. \citealt{2021Jin} or \citealt{2020Sharon}).
The external \nifs\ component is placed from $f_2$, to $f_3$, so it can be truncated below the surface, or reach it if $f_3=1$.
The total mass of \nifs~ varies as result of the change of any $X$ or $f$ parameter, therefore, we decided to indicate its value in parentheses in the figures presenting the LC that result from the individual parameter variation. In this section, we discuss the morphology of the first and second peaks, but without mentioning the earlier initial spike due to the shock breakout, since it is much brighter and faster than the peaks discussed here.

\begin{figure*}
\resizebox{\hsize}{!}{\includegraphics[width=0.45\hsize,keepaspectratio]{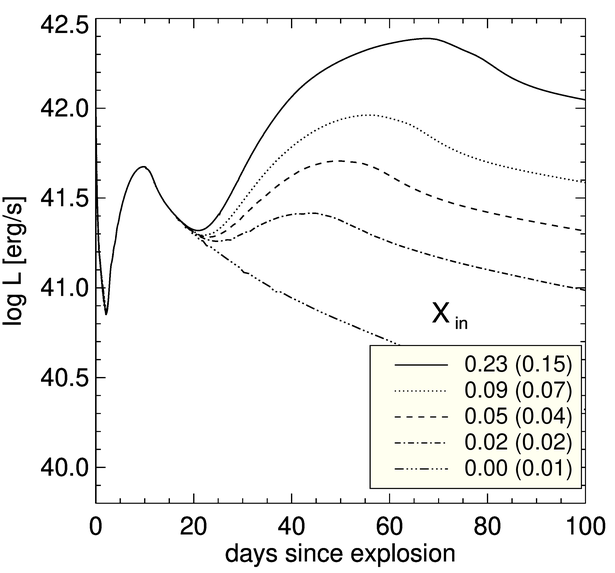}%
\includegraphics[width=0.45\hsize,keepaspectratio]{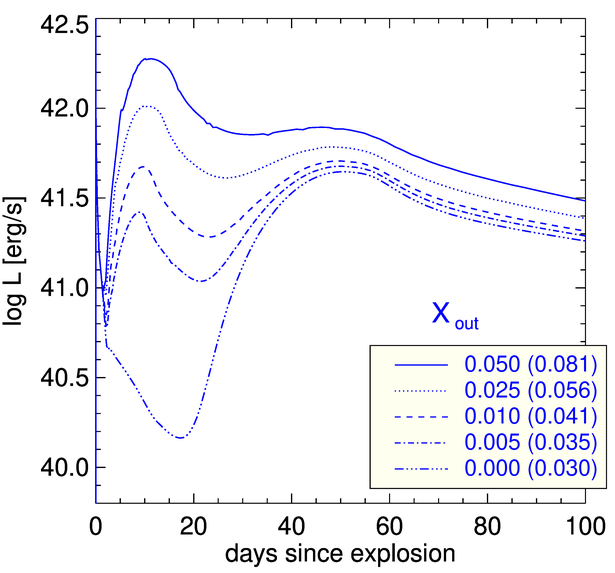}}
\caption{
Effect of the variation in the abundance of \nifs , with the other parameters fixed.
All these calculations correspond to the {\tt He11} progenitor, with $E_{\rm exp}=2.0$~foe, a $2.15$~\Msun\ compact object (then $f_0=0.199$), $f_1=0.25$, $f_2=0.91$ and $f_3=0.991$. Total \Nimass ~given in units of solar masses is indicated in parentheses. 
Left: Effect of variation of the inner abundance $X_{\rm in}$. Right: Same effect but related to the external abundance $X_{\rm out}$. In the left panel, the outer abundance of \nifs ~is set to $X_{\rm out}=0.01$ and  $X_{\rm in}= 0.047$ in the right panel.}
\label{fig:Xi-var}
\end{figure*}

\begin{figure}
\resizebox{\hsize}{!}{\includegraphics{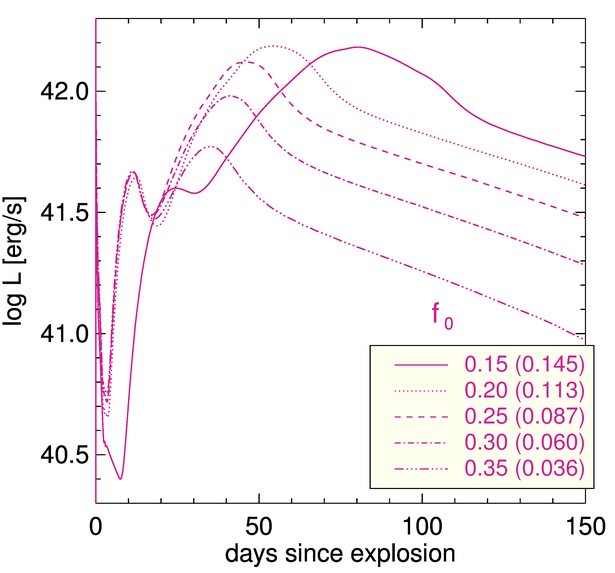}}
\caption{Effect of the variation in the parameter $f_0$ that, for the fixed progenitor (here {\tt He11}), defines both the mass of the ejecta and of the compact object. The numbers in parentheses are the total \Nimass\ in \Msun.
In this case $f_1=0.4,\, f_2=0.85,\, f_3=0.95,\, X_{\rm in}=0.047,\, X_{\rm out}=0.01$.
} 
\label{fig:f0-var}
\end{figure}

\begin{figure*}
\resizebox{\hsize}{!}{\includegraphics{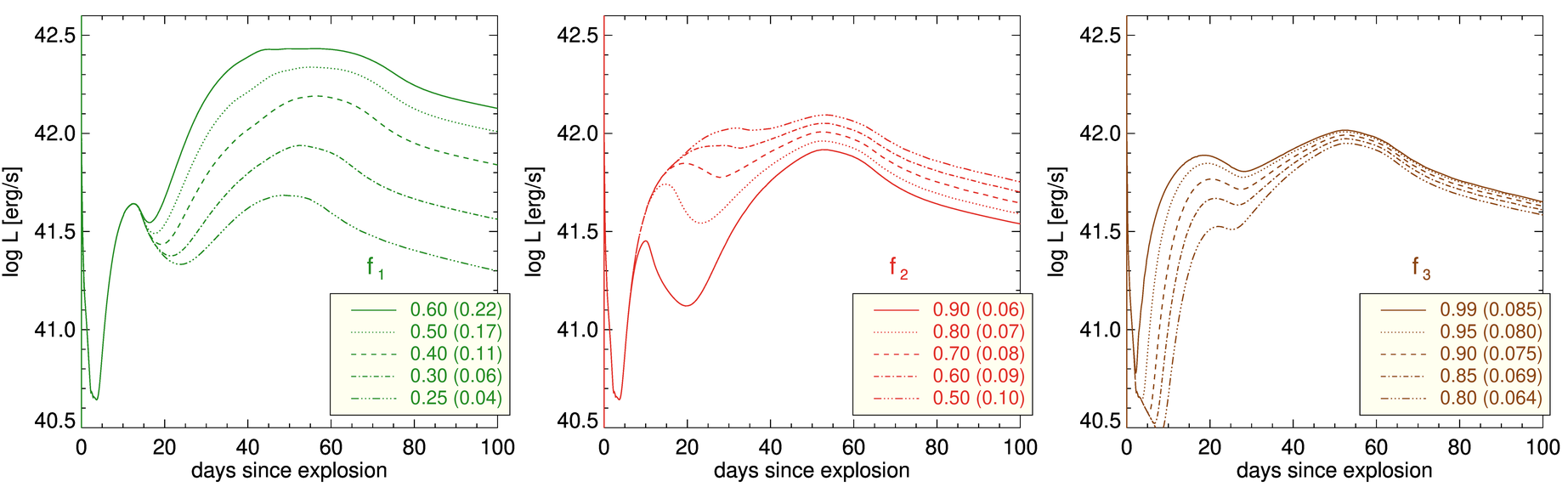}}
\caption{Variation of the fractions $f_i$ for the {\tt He11} progenitor, with $E_{\rm exp}=2$~foe, a fixed $2.15$~\Msun\ compact object (then $f_0=0.20$), abundances $X_{\rm in}=0.05$ and $X_{\rm out}=0.01$, and the varying parameter indicated in each panel. 
For the left panel: $f_2=0.85$ and $f_3=0.95$.
For the middle panel: $f_1=0.30$ and $f_3=0.95$.
In the rigth panel: $f_1=0.30$ and $f_2=0.70$.
With the abundances fixed, the mass of \nifs\ change according to the variation of the distribution parameters. The total \Nimass\ is indicated in parentheses. 
} 
\label{fig:fs-var}
\end{figure*}

Figures~\ref{fig:Xi-var}, \ref{fig:f0-var}, and \ref{fig:fs-var} show the effect on the bolometric LC of the variation of each of the parameters that define our nickel distribution, keeping the others fixed.
For these calculations, we used {\tt He11} as the initial configuration, with explosion energy of 2~foe (foe$\equiv$ 10$^{51}$ erg) and a compact remnant of 2.15~\Msun\ (except in Fig. \ref{fig:f0-var}, where we allow for a variation of the remnant mass). A variety of two peaked LC are produced, ranging from curves with two clearly separate peaks to cases where both peaks begin to merge. The luminosity of the first (second) peak are mainly dominated by the external (internal) abundance, although the extension of the nickel-rich region also has an important effect. We note that when $ f_1 $ is adjusted closer to $f_2$, the peaks blend into a more subtle double-hump morphology. The timescale of the first rise is also related to $f_3$ that controls the outermost border of the \nifs\ distribution. 
As found by \cite{2017Noebauer}, the rising phase duration is related to the optical depth of the material placed on top of the border that we denote $f_3$. The rate of $L$ decline after the first peak depends on $f_2$ (as can be see in Figure~\ref{fig:fs-var}). 

In a previous analysis performed by \cite{2021BAAA}, we study a double-nickel distribution for a less massive progenitor, ({\tt He4}). From that analysis, together with the results presented here, we found that a
large temporal departure between the light curve maxima is a consequence of depositing the inner and outer nickel at a larger distance in mass coordinate, that is, if $f_2$ is much greater than $f_1$.
Furthermore, the temporal separation of the peaks seems more prominent for a larger ejected mass, produced for a more massive progenitor. For a fixed progenitor, the mass of the ejecta is controlled by $f_0$ and the effect of its variation can be seen in Figure~\ref{fig:f0-var}.

\begin{figure}
\resizebox{\hsize}{!}{\includegraphics{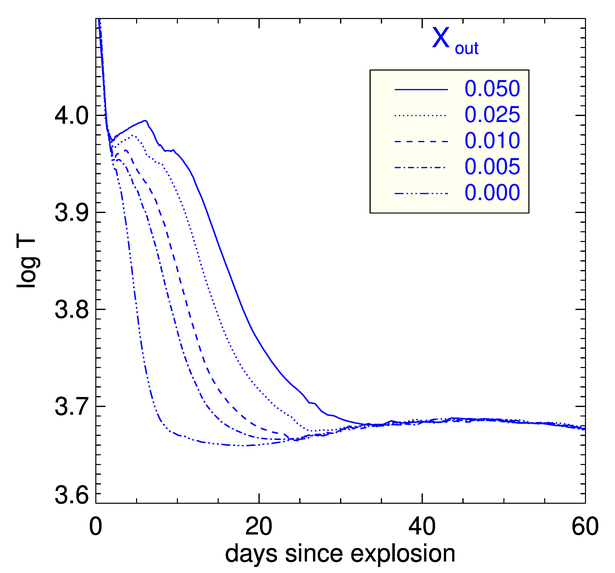}}\\
\resizebox{\hsize}{!}{\includegraphics{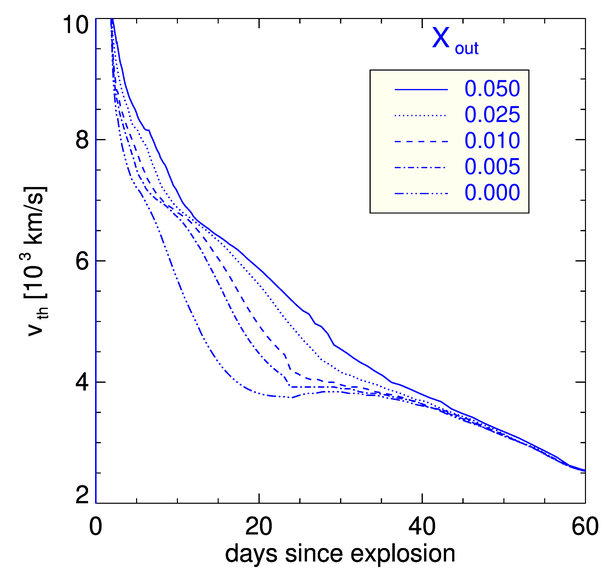}}
\caption{Effect of the external abundance of nickel on the temperature and velocity evolution. The other parameters have the same values as in the right panel of Figure~\ref{fig:Xi-var}. The curves show a zoom on the temperature and velocity at the color depth (also known as the thermalization depth) before the second peak of the LC. Afterwards, they all behave similar to the case $X_{\rm out}=0$.} 
\label{fig:Xe-thermal}
\end{figure}

The effect of the $X_{\rm out}$ variation is presented in Figure~\ref{fig:Xi-var}. It has a clear correspondence with the luminosity of the first peak in the LC. We show also an accompanying plot of the evolution of the temperature and velocity at the thermalization depth (Fig~\ref{fig:Xe-thermal}). 
As suggested by \cite{EnsmanBurrows1992}, the 'thermalization' depth can be estimated as the layer where
the continuum is actually formed, given by the condition
$\tau_{\rm abs} \tau_{\rm sct} \sim 1$,  where $\tau_{\rm sct}$ is the optical depth for scattering and $\tau_{\rm abs}$ is the optical depth for absorption.

The first rise in the LC after the shock breakout results from the heating power provided by the radioactive decay and this seems to be a distinctive feature of the scenario, although it is subtle if the external \Nimass\ is small. In the first few days (around day five since the explosion), the temperature $T_{\rm th}$ may even rise for a while before decreasing again. An increase in temperature should have an effect on the broadband photometry,
producing a differential change in flux depending on the wavelength. This is probably attributable to a color effect, although this change may be slight, especially if the external mass of nickel is small.
If such a behaviour were to be found in a given SN along with the first peak in the LC, this is an strong sign of an extended distribution of radioactive material in the outer ejecta regions, as in the scenario we are studying here. As part of the energy transforms to kinetic, it is also accompanied by an increase in velocity. Both panels include the case without external \nifs\ for comparison. 

\section{Published double-peaked SNe}
\label{sec:observadas}
\begin{figure*}
\resizebox{\hsize}{!}{\includegraphics{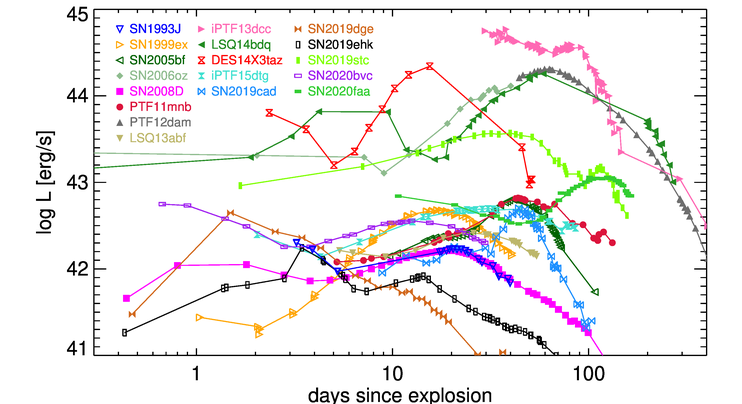}}
\caption{Data from the literature, as detailed in Table~\ref{LCobs}, displays a variety of double-peaked SNe with early detections. For visualization, the error bars are not included. For objects that present a clear double-peaked LC, we see that the morphology in both peaks is quite diverse and it is likely that the physical explanation of these is varied as well. 
Some SLSNe are also shown and, in addition, PTF12dam, which is not double-peaked, is included as a reference of the SLSN.} 
\label{fig:obs1}
\end{figure*}

In recent years, thanks to great observational efforts, the number of SNe that present an emission prior to the main maximum of their light curve has notably increased. Currently, this early emission has been detected in almost all type of SNe, although the detection frequency is dependent on the type of supernovae, which may be related to their origin.
The list in Table~\ref{LCobs} is a collection of reported SNe, including a few superluminous SNe that have shown early emission. 
They illustrate the underlying diversity in how the designation of double-peaked SNe is applied.
Figure~\ref{fig:obs1} shows the data set from this list of references and allows us to make some comparisons. First, we note that there is a wide range of luminosities, timescales, and, hence, slopes in the LCs. 
Our intention has been to remain thorough with regard to the population of normal SNe having an initial rise followed by a double peak in the observed bolometric data. 

In this selection, we excluded all the SNe that manifest only an excess of emission in the bolometric LC, despite of having two distinct
light curve components in one or more  photometric bands. Sometimes that excess is only modest and noticeable when one focus on the luminosity data at epochs around the main peak. For example, SN2013ge displays such a morphology, mentioned as a shoulder in the LC description \citep{2016Drout}.
Interestingly, in order to explain this case, the authors have considered a model with outwardly mixed \nifs\ and incomplete gamma-ray trapping to explain the rapid decline.

Type IIb SN1993J has attracted great attention (e.g. \citealt{1994Chevalier, Woosley1994, Nakar2014}, and references) and it has been studied for many years and in different wavelengths (in radio, e.g. \citealt{2009Marcaide,2011Marti}; and most recently published in IR by \citealt{2022Zsiros}).
Among the few SNe of type Ib included in the list, we aim to dedicate more attention to SN2008D (in Section~\ref{08D}).
The SLSN2006oz presents a plateau-like morphology, with a duration of 6--10 days before the rise of the main peak, which is more evident when the axis of time has linear scale. \cite{2016quark} set up a specific scenario to explain the origins of this LC, which consists of the transition from a neutron star to a quark star. The more conventional optional models are not well constrained in this case, due to the lack of post-maximum observations \citep{2012Leloudas}.
We refer to \cite{2016Smith} for the SLSNe that have well-measured pre-explosion photometry, some of them with double peaks. Next, we  focus onto the SNe that that have not been classified as superluminous \footnote{We do not simply refer to a 'normal' supernovae because SN2005bf is somewhat brighter than the mean SNe.}.

A different but fascinating case (albeit excluded from our list) is offered by the hydrogen-rich SN2009ip and its analog SN2010mc \cite{2014Margutti}. Their status of true SNe has been questioned, suggesting that they may represent  another type of violent, non-terminal event.

The calcium-strong transients, such as SN2019ehk or SN2021gno, are SNe that present atypically high calcium-to-oxygen nebular line ratios. They have generated a consistently inconclusive debate with regard the preferred scenario \citep[see][and references therein]{2021Jacobson,2019Shen,2022Jacobson}; thus, the origin of these events is unclear.
It is worth mentioning that these cases show a fast evolution of the early bolometric LC and their modelling with a double \nifs\ profile could be comparable to the case we discuss in Subsection~\ref{08D}.

\begin{table*}
\begin{center}
\caption{Light curve references. When not stated, there is a reasonably accurate estimate of the time of explosion, $t_0$.}
\begin{tabular}{lcll}
\hline
\hline
\noalign{\smallskip}
Name & Type & Reference   &Comment     \\
\noalign{\smallskip}
\hline
\noalign{\smallskip}
SN1993J   & IIb   & \cite{1993Ray} &\\
SN1999ex  & Ib/c  & \cite{2002Stritzinger} & \\
SN2005bf  & Ib/c  & \cite{Folatelli06} & \\
SN2006oz  & SLSN I& \cite{2012Leloudas} & days since discovery\\
SN2008D   & Ib    & \cite{2009Modjaz} & XRT080109\\ 
          &       & \cite{2013Bersten} & earliest $L_{\rm bol}$ data\\
          &       & \cite{2009Tanaka_b}& latest data\\
PTF11mnb  & Ic    & \cite{Taddia18} & Stripped envelope\\
PTF12dam  & SLSN I& \cite{2017Vreeswijk} & $t_0=70$ d, slow decline, no double\\ 
LSQ13abf  & Ib    & \cite{2020Stritzinger} & \\
iPTF13dcc & SLSN I& \cite{2017Vreeswijk} & $t_0=89$ d\\
LSQ14bdq  & SLSN Ic& \cite{2015Nicholla} & \\
DES14X3taz& SLSN I& \cite{2016Smith} & \\
iPTF14gqr & Ic    & \cite{2018De} & SN 2014ft\\ 
iPTF15dtg & Ic    & \cite{2016Taddia} & \\
SN2019cad & Ic    & \cite{Gutierrez2021} & err $\Delta t_0=4$ d\\
SN2019dge & Ib    & \cite{2020Yao} & first data with large error\\
SN2019ehk & Ib    &\cite{2020Jacobson, 2021Jacobson}& Ca-rich. err $\Delta t_0=0.1$ d\\
          & IIb   &\cite{2021De}& \\
SN2019stc & SLSN I& \cite{Gomez2021} & \\ 
SN2020bvc & Ic-BL & \cite{2020Ho} & similar to SN2006aj, X-ray detection\\
SN2020faa & SLSNII & \cite{2021Yang}& iPTF14hls-like\\ 
\hline
\noalign{\smallskip}
\hline
\end{tabular}
\label{LCobs}
\end{center}
\end{table*}

Figure~\ref{fig:obs1} shows that the SNe that has early detections are more often discovered as they pass through a decline in the bolometric luminosity. This pattern does not necessarily exclude a previous rise.  It is then convenient to clarify what we mean for the early phases when describing the observational $L_{\rm bol}$ data of double-peaked SNe. The early phase consists of the epochs between the discovery and the first time the luminosity trend changes, either at a local minimum (so there was an early decline only detected) or at a local maximum (an early rise was observed). When a given SN exhibits an early decline, there are two possibilities: either we are seeing the end of the cooling phase or this decline follows the end of a non-detected first maximum.


\subsection{A selected family}\label{family}

\begin{figure}
\resizebox{\hsize}{!}{\includegraphics{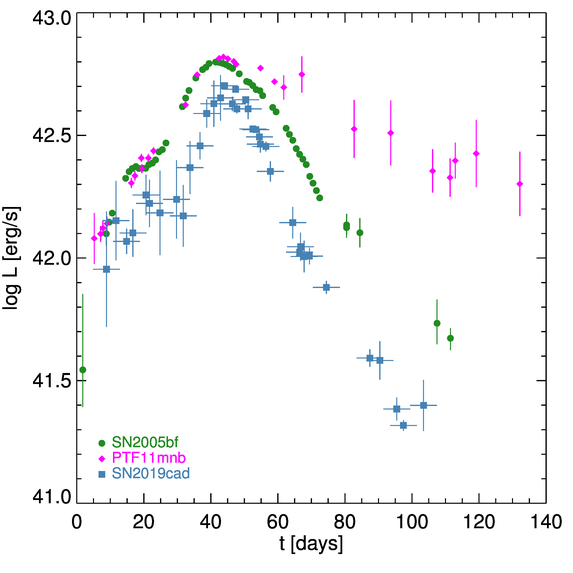}}
\caption{Three objects selected to apply the proposed double-peaked model. The presence of a raising phase before the first peak and the time of this peak were a critical characteristic for our sample selection. We also note that the time of the second peak is at an epoch much later that what has typically been observed for SESNe.}
\label{selected}
\end{figure}

From all the events shown in Figure~\ref{fig:obs1}, we selected a group that share the observational characteristics of the two maxima clearly present in the bolometric LC with an initial rise in luminosity prior to the first maximum. 
We restricted our study to objects that have not been classified as SLSNe. 
We are left with the few cases that show analogies in the LC to SN2005bf. This SN has very accurate measurements, except for the first and the very late ($t >70$~d) bolometric points \citep{Folatelli06}. 
There are other two SNe that clearly resemble SN2005bf; PTF11mnb \citep{Taddia18} and SN2019cad \citep{Gutierrez2021}. For the latter, we considered the zero--extinction data. These three LCs can be compared in more detail in Figure~\ref{selected}, which  includes the available error bars. In each case, a double \nifs\ profile was considered in order to model the LC.
Our new contribution is to use a unique function for the nickel abundance profile $X (f)$, the one  presented in Section~\ref{sect:models}.

\begin{figure*}
\resizebox{\hsize}{!}{\includegraphics{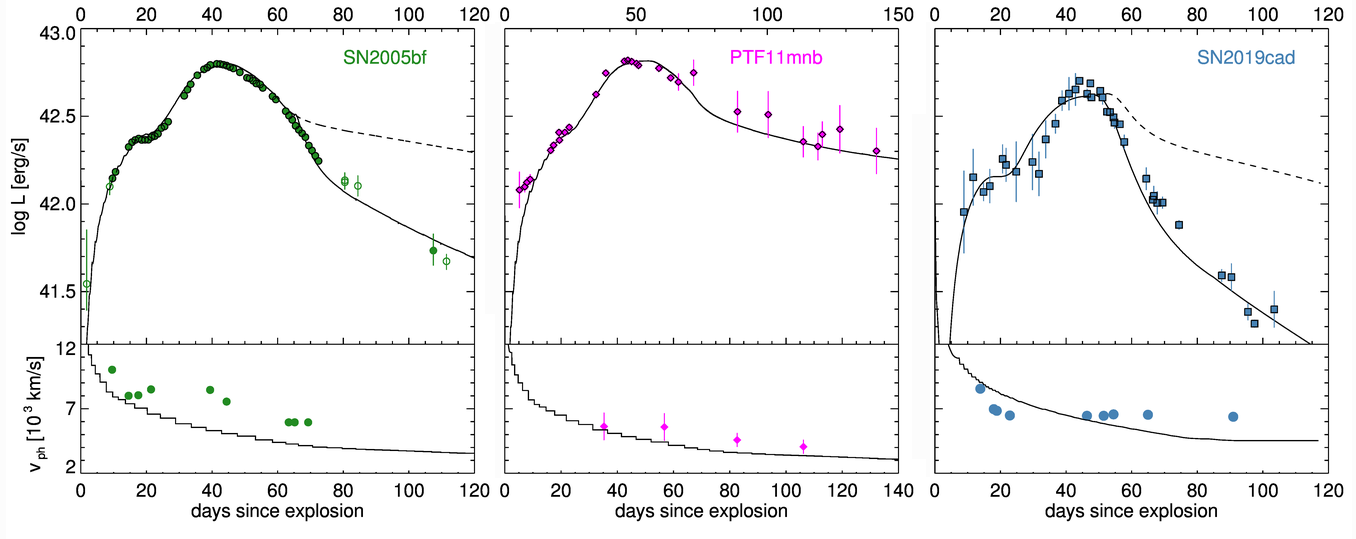}}
\caption{Fitting models for our selected sample, SN2005bf, PTF11mnb, and SN2019cad, within our parametric prescription for the \nifs\ double distribution profile. The late light curve decline slope for SN2005bf and SN2019cad is reproduced when passing from a fixed $\kappa_\gamma$ opacity (dashed line) to one adjusted to account for the gamma-ray leakage (solid line).
}  
\label{fits_LC}
\end{figure*}

Table~\ref{tabla_parameters} presents our fitting parameters for this selected family of SNe.
The resulting LC are shown in Figure~\ref{fits_LC}. 
We also show a panel with the velocity evolution, but we notice the models are not meant to explain photospheric velocities. However, the comparisons with velocities give reasonably good results, especially for PTF11mn and SN2019cad.

For SN2005bf and SN 2019cad, the hypothesis of a constant standard gamma-ray opacity does not accommodate the complete observed LC, as was found in the aforementioned references. A modified leakage is invoked at late times. Our hydrodynamic code is not well designed for the phases with optically thin conditions, this fact could help explain the need for an adaptation of $\kappa_\gamma$values.  

According to our results, for SN2005bf and SN2019cad, the turn-down of the gamma-ray opacity required to match the fast decline in the LC after the main peak is a decrease of $\kappa_\gamma$ by a factor of 10 and 60, respectively. 
In support of this hypothesis, we note that \cite{2009Tanaka_05bf} found polarimetric evidence for an aspherical, possibly unipolar, explosion in the case of SN2005bf. 
A significant change of the gamma-ray leakage could be expected for SN2019cad given the LC analogy with SN2005bf.

Instead, PTF11mnb does not require an adaptation with reduced gamma-ray opacity, but we note the large error bars in the bolometric data at epochs after $\sim 60$ days.
In this case, the alternative scenario provided by \cite{Taddia18} consist of a hybrid combination with a smaller \Nimass $\approx 0.11$ \Msun\ from Arnett model, plus a magnetar with $B=5\times 10^{14}$~G and $P=18.1$~ms. However, the authors disfavour this magnetar because nickel stands to offer a simpler explanation for the overall LC. 

A similar proposal has been discussed for SN2005bf \citep{2007Maeda}, but with a relatively large braking index as an alternative to full energy trapping. A magnetar also works for SN2019cad \citep{Gutierrez2021} with the \nifs\ already accommodated into a two-zone profile. We note that the formation of a magnetar and the eventual presence of outflows that are thought to reshape the nickel profile could, in principle, coexist \citep{2021Shankar}.

The need to tune the value of $\kappa_\gamma$ can be thought as a drawback for the double \nifs\ model, but it can be roughly justified in relation to the asymmetries of the explosion. 
For a sample of normal SNe, \cite{2015Wheeler} found very different late-time tails and they analysed the source of their heterogeneous slopes in relation to the trapping of gamma-ray energy from radioactive decay.
It is interesting to mention that although 
\citep{Chen2015} postulated (in another context) that to model the late epochs ($t>200$~d) of PTF12dam, a modest increase in the escape of gamma-ray radiation was needed. We note that for the magnetar scenario, the energies of the gamma-ray photons can be quite different than those from the radioactive decay, therefore entering into a range where it is valid to suggest that the opacity for the high-energy radiation is dominated by pair production and becomes one order of magnitude lower than in previous phases. A similar treatment of the late-time leakage of hard radiation was invoked by other authors (e.g. \citealt{2015Wang, 2017Nicholl, 2022Cartier}); whereas, recently, a more consistent approach was investigated by \cite{2021magnetar}.
 

For SN2019cad, we show a model very similar than previously obtained in \cite{Gutierrez2021}. In the present case, the inner \nifs\ component has a flat abundance, while in the previous study, $X_\text{in}$ exhibited a slight outward increase with mass coordinates to mimic the abundances of Si-Ar-Ca. 
Here, the LC shape of SN2019cad is reproduced by a pre-SN mass of 11 \Msun, and an explosion energy of 4~foe, which is slightly greater than the 3.5~foe reported in \cite{Gutierrez2021}.
The double-peaked \nifs\ distribution has an external component of 0.049 \Msun and an internal component of 0.28 \Msun. The LC after the main peak drops steeply. To model this effect, an enhancement of the gamma-ray leakage is assumed at around the date of $L$-main peak: a constant $\kappa_{\gamma}=0.03$ is switched to  $\kappa_\gamma=0.001$ cm$^2$ g$^{-1}$. As in the model presented by \cite{Gutierrez2021}, the external nickel needs to be close but below the surface to fit the time scale of the first peak of the light curve. 
For the cases of SN2005bf and PTF2011mnb, it is not straightforward to establish a detailed comparison between our results and the models of \cite{Folatelli06} and \cite{Taddia18} because of the different \nifs\ profile. 
In \cite{Taddia18}, the nickel profile for PTF11mnb is scaled from SN2005bf.
The published values of total \Nimass\ are 0.6 and 0.59 \Msun,~respectively, while we obtained 0.448 \Msun\ for SN2005bf and  0.499 \Msun\ for PTF11mnb.

\begin{table*}
\begin{center}
\caption{Model parameters}\label{tabla_parameters}
\begin{tabular}{cccc}
\hline
\hline
\noalign{\smallskip}
Parameter             & SN2005bf                     & PTF11mnb                  &SN2019cad $^1$\\ 
\noalign{\smallskip}
\hline
\noalign{\smallskip}
$M_{\rm ej}$           &   $6.1$ M$_\odot$            &  $6.1$ M$_\odot$          &   $9.55$ M$_\odot$\\
$M_{\rm preSN}$        &   $8$ M$_\odot$              &  $8$ M$_\odot$            &   $11$ M$_\odot$\\

$E_{\rm exp}$                  &   $1.7\times 10^{51}$ erg    &  $1.5\times 10^{51}$ erg  &   $4\times 10^{51}$ erg\\


$\kappa_\gamma$ [cm$^2$/g]& 0.03 $t\leq 65$~d           &  0.03 all epochs          &   0.03 $t\leq 45$~d\\ 
                        & 0.0018 $t> 65$~d            &                           &   0.001 $t> 45$~d \\

\noalign{\smallskip}                     
\hline
\noalign{\smallskip}
$f_0,\,f_1$            & 0.2, 0.247                   & 0.2, 0.259                &   0.13, 0.154             \\
            $f_2,\,f_3$& 0.524, 0.99       &             0.563, 1                 &   0.646, 0.947\\

$X_\text{ in}$,
$X_\text{ out}$         & 0.952, 0.029                & 0.960, 0.029              &   0.99, 0.015\\
\Nimass\  in, out $^a$   & 0.352, 0.096 \Msun          & 0.395, 0.104  \Msun        &   0.283, 0.049 \Msun\\

\Nimass\  total   & 0.448 \Msun          & 0.499  \Msun        &   0.332 \Msun\\

\noalign{\smallskip}
\hline
\hline
\end{tabular}
\begin{list}{}{} 
\item $^a$ \footnotesize{\Nimass\ is afterwards computed, not an initial parameter.}
\item $^1$ \footnotesize{For the data with host-galaxy $E(U-B)=0$ in \cite{Gutierrez2021}.}\\
\end{list}
\end{center}
\end{table*}

\subsection{An extreme example}
\label{08D}

\begin{figure}[h]
\resizebox{\hsize}{!}{\includegraphics{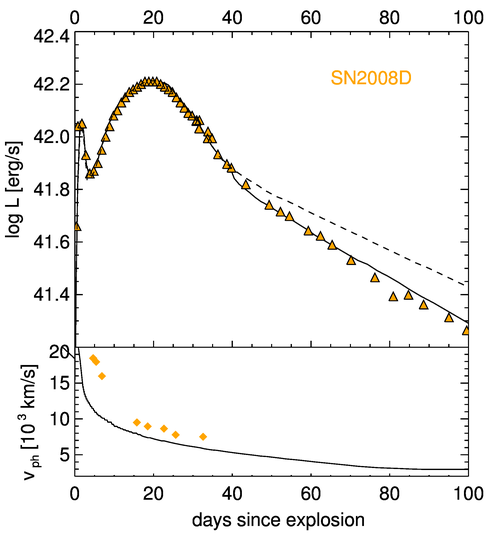}}
\caption{SN2008D requires a strongly inverted \nifs~ profile, with the external abundance of $X_{\rm out}=0.71$ in a layer that reaches the surface; in this case, the progenitor model is {\tt He5} with $E_{\rm exp}=1$~foe. The late LC decline is reproduced when passing from a fixed $\kappa_\gamma=0.03$~cm$^2/g$ opacity (dashed line) to half this value at $t>40$~d (solid line).
The lower panel shows the evolution of the photospheric velocity, although we did not use observational velocities to constrain the model. The velocities of the photosphere are taken from \cite{2009Tanaka_08D}. 
}
\label{SN2008D}
\end{figure}

A very interesting feature noted for SN 2008D was the early  detection in X-rays and multiband follow-up \citep{2009Modjaz}, whose combined interpretation has been debated in the past (see also \citealt{2017Branch} for a summary of the open issues).
In Figure~\ref{SN2008D}, we show our modelling of SN2008D light curve. This SN has a fast evolution in the LC with the main peak (the second) at a timescale of $\sim 17$~d. We note this is similar to the time when the SNe of the sample of Section~\ref{family} present the first and weakest peak. That early peak can be therefore interpreted as indicative for a \nifs\ profile with rather different values of the parameters than those treated in the previous section, which makes this SN an intriguing extreme object.

This morphology is particularly challenging for other models, since a more promising case is represented by the explosion of a Wolf Rayet star through a thick wind (see discussions by \citealt{2018Dessart,2011Rabinak}, and references).
Models with an envelope give a reasonable fit to the early observations with the exception of the earliest data point, that was considered in \cite{2013Bersten} analysis, but not in other LC models for SN2008D. For the main peak, \cite{2009Tanaka_08D}  discussed the degeneracy of the parameters between the possible progenitors, with the \Nimass $\sim 0.05 - 0.07$~\Msun and the \nifs\ efficiently mixed into the outer ejecta layers.

After considering different envelopes and thick wind configurations to fit all of the early SN2008D data, \cite{2013Bersten} assumed a double-nickel distribution as an alternative and found that it improved the agreement with the data.
Here, we considered, for SN2008D, a progenitor star with main sequence mass of 18~$M_{\odot}$,  producing, at the moment of the explosion, a Helium core of $\sim 5$ \Msun, denoted by {\tt He5}. The evolution of this progenitor model was calculated by \cite{1988Ken}. 
To fit the first peak, the double profile of nickel is pushed to the extreme case where it reaches the surface of the star and the abundance of radioactive material is particularly high there ($X_{\rm out}=0.71$), a value that is probably unphysical. The model shown here has \Nimass\ of 0.074 \Msun\ in the inner region and 0.018~\Msun\ at the outer layer, that is larger than 0.01~\Msun\ for the external \Nimass\ in \cite{2013Bersten}, but we note that also the exploded progenitor star is different ({\tt He8} in their case). 
In our scheme of parameters, the internal component extends from $f_0=0.36$ up to $f_1=0.93,$ with $X_{\rm in}=0.026;$ while the outer one goes from  $f_2=0.995$ to $f_3=1$.

We simulated the explosion with a typical value $E_{\rm exp}=1$~foe, rather closer to the energy in the simple model from \cite{2008Chevalier}. In particular, SN2008D does show a less flat velocity evolution than the SNe in the previous section.
If the velocity data were considered when building the model, it is likely that a more energetic explosion would be favoured to explain the fast ejecta, as found by \cite{2013Bersten}. 
Similarly to our approach to modelling SN 2005bf and SN 2019cad, a change in the trapping of the gamma-rays, is needed to improve the match with the LC at late epochs ($t>40$~d). Specifically, we introduce a slight change, lowering  the value of $\kappa_\gamma$ by a factor of 2. 

This model is still simple and we cannot make further comments on the spectroscopic features. \cite{2018Dessart} modelled SN2008D prior to the re-brightening as the explosion of a lower mass (2.73 \Msun) star with an extended and tenuous envelope. 
The authors most vocal critique with regard to the external \nifs\ is that it should amplify non-thermal effects in the outer layers, generating very broad He{\sc I} absorption lines at one week -- which were not found in the observational data. 
However, these authors were unable to properly model all of the existing early data.
The newest models for the type Ic supernovae LSQ13abf and SN 2008D are presented by \cite{2021Woosley}, including a consideration of their progenitor evolution in close binaries.

\section{Discussion and conclusions}\label{sec:discuss}

The possibility of using an arbitrary distribution of \nifs\ in our code offers the chance to investigate the effects of any spherically symmetric profile. 
In the scheme we considered in this work, the alleged presence of jet-like outflows in the collapse of a massive star could drive the presence of external radioactive material that could well be responsible for the doubled distribution of \nifs. 
A more consistent justification of this scenario would emerge from a thorough understanding of the explosion mechanism itself and the development of calculations that (following the propagation of the outflows) could properly locate the nucleosynthetic yields of radioactive elements into the stellar interior.
Studies in this direction were made by \cite{1999Kok}, while a discussion on the radioactive isotope yields from various jet-like explosions of magnetorotational core-collapse SNe was initiated by \cite{2015Nishimura}. 
The case of a jet choked inside of the star is of particular interest. Related studies have been presented by \cite{2015Nakar} and \cite{2016Senno}. In addition, an analysis by \cite{2017Soker}, focussed on SLSNe, called for a paradigm shift from neutrino-driven to jet-induced explosion models of all core collapse supernovae.

The calculated LCs occur with different timescales depending on several parameters beyond the radioactive content -- be it stellar- or explosion-related. 
The strongest conclusion from our results regards the SNe whose data resemble the SN2005bf morphology, namely, those with an early rise before the first peak of the LC. In this scenario, the first peak after shock breakout in the bolometric LC would be potentiated by the outer \nifs. This is reasonable since the energy coming from the outer radioactive material is covered by an optically less thick layer, then its energy emerges sooner than the outcome from the inner \nifs. ~Instead of accommodating the outer nickel as part of one distribution strongly mixed outwards and a different power source for the main peak, we have two separated layers that are radioactively enriched. 
We re-estimated the nucleosynthetic yields within this prescription for the selected sample of reported SNe that are SN2005bf-like, plus the case of SN2008D, which has less available data on an early rise. For SN2008D, that early rise was only once modelled within this scenario by \cite{2013Bersten}.
If the model depicted here is considered acceptable for the case of SN2008D, it could be noted that it would thus exhibit one of the shortest timescales for the first peak. This encourages the proposal that the model can also accommodate other SNe as well. However, the contrast in luminosity at the earliest phases is also important. 
We cannot place a strong statement on the remaining SNe shown in Section \ref{sec:observadas} because of all the other variables that need to be further explored and that are beyond the scope of the current work.

We have designed a systematic study to explore the applicability of a simple \nifs\ distribution to model some of the double-peak SNe. We have considered that the outer mass of nickel is in all cases lower than the inner one. 
This is regulated through the abundance and extent in mass coordinate. 
We considered a fixed H-poor progenitor ({\tt He11}) here, whereas for a less massive case, we previously presented a brief study in \cite{2021BAAA}. Our prior results are consistent with the exploration of the parameter space presented here. Based on a direct comparison, the less massive star produces a more rapid evolution of the overall LC, but a simple relation between the timescales for the peaks and the \nifs\ profile does not appear to be straightforward.

The subsequent decline after the first peak does not seem entirely consistent with the rapid early decline observed in many SNe.
If the first luminosity data fall very steeply, this is not what we would expect from nickel power, unless it has an extreme distribution. Therefore, the observed early decline slope of the LC could minimize doubts about a prior, non-detected rise.
For a given SN, this comparison is valid if the explosion date is well determined, however, because of the inherent difficulties of the discovery at the earliest stages after the explosion, sometimes this information is not available. 
We cannot rule out that our selection of SNe with an initial bolometric rise may suffer from an observational bias that makes the sample so small. 

Diffusing the radioactive energy out of the ejecta is a mechanism that combined with the radioactive decay timescale cannot produce an arbitrarily early first peak. Our exploration indicates that around one day offers only a limiting case.
In the extreme distribution, we fit to the LC of SN2008D, the first peak happens at around  $\sim 1.3$~days as a consequence of the external nickel located in a very thin, superficial layer, with very high \nifs\ abundance.
\cite{2017Noebauer} explored the effect of other short-living isotopes that could be important into the shaping of the early phases of the LC. That additional energy source, provided by $^{52}$Fe and $^{48}$Cr decay channels, is not yet implemented in our code.

In the cases where the outer \nifs\ is deeper and less concentrated, the rise takes few days since the explosion and, therefore, the first peak is produced later. This is related to the decay half-life of the radioactive elements we are considering.
Moreover, if the first peak occurs within $\sim 10 -20$ days after the explosion epoch as in the case of SN2005bf-like objects, the outer abundance found is moderate ($\sim 0.1$ at most) and the radial extent in mass is wider (that is, $f_2$ lower). 
We note that other models face difficulties in producing an early rise with the first peak within this timescale, which is coincidental with the time when other SNe exhibit their maximum (see for instance \citealt{2015Gaitan}). 
In the cases we investigate here, well-covered photometry prior to the first peak was helpful in constraining how deeply the external \nifs-rich material should be placed. The radioactive \nifs\ that powers the second peak of these objects is concentrated close to the compact object.

The so-called hybrid models containing different sources of energy can work for many SNe.
The question of whether this model, coupled with a doubled distribution of \nifs,\ is a good alternative seems to be related to the first slope of the observed bolometric LC, with an early rise preferred over an early decline.
The timescale of the minimum between the peaks is somehow related to the progenitor mass; however, to put quantitative constrains, the whole parameter space should be further explored, namely, the nickel profile  varied in combination with other physical quantities that we have considered to be fixed here (the explosion energy, $E_{\rm exp}$, the stellar characteristics such as mass and radius, as well as the leakage capacity for the gamma rays or other transport parameters). Any improvement on the model should be also accompanied by the inclusion of the velocity data into the estimation of the goodness of the fit (see discussion in \citealt{Martinez2020} and references therein), which we have not done here.

In order to explain the observed late-time slope of the LC, we had to  abandon the idea of the complete trapping of gamma-rays and thermalization of the energy deposited in the ejecta. Instead, a few days after the second maximum, this was accounted by introducing a change in $\kappa_\gamma$. This may emulate the fact that even if a gamma-ray photon is absorbed, a fraction of its energy does not thermalize and is allowed to leak out.
This change is phenomenologically motivated and case-dependent. We cannot provide a rigorous quantitative justification of this assumption. However, this is part of a long-known problem that produces great heterogeneity in the late decline in the LC (\citealt{1997Clocchiatti}, and references).
A variable gamma-opacity has been suggested to result from asymmetries in the ejecta \citep{2017Branch} 
and the gamma-ray leakage could also be enhanced due to the presence of small-scale inhomogeneities, for instance, low-density holes or clumps in the ejecta that we cannot catch in our 1D code.
We note that 3D calculations with the pertinent details have been developed by \cite{2020Jerkstrand} for the modelling of decay gamma-ray photons that are not captured by the ejecta. Their detection points to a large degree of \nifs -mixing and confirms that some of the high-energy photons leak out.
In an independent way, the gamma-ray transport was simulated through Monte Carlo techniques by \cite{2019Wilk}, who suggested a time and spatially varying gray opacity factor could improve results within the approximation of \cite{1995Sutherland}, namely, the one we apply in this work.

Nowadays, early detections of SNe are becoming more frequent and there are many cases where the detected emission poses a challenge to existing models. Therefore, it is plausible that more exotic explanations need to be explored to explain these LCs. In this sense, calculations such as those presented in this study can help elucidate the underlying physical scenario.


\begin{acknowledgements}
We thank the anonymous referee for the comments and suggestions that have helped to improve the paper.
This research was partially fund by UNRN PI2020 40B885,  and grant PICT-2020-SERIEA-01141. We also acknowledge the support of CONICET through project PIP 112-202001-10034.
\end{acknowledgements}

\bibliographystyle{aa}
\bibliography{refs}

\begin{thebibliography}{87}
\expandafter\ifx\csname natexlab\endcsname\relax\def\natexlab#1{#1}\fi

\bibitem[{{Anderson}(2019)}]{Anderson2019}
{Anderson}, J.~P. 2019, \aap, 628, A7

\bibitem[{{Banerjee} \& {Mukhopadhyay}(2013)}]{Banerjee13}
{Banerjee}, I. \& {Mukhopadhyay}, B. 2013, \apj, 778, 8

\bibitem[{{Bersten} {et~al.}(2011){Bersten}, {Benvenuto}, \&
  {Hamuy}}]{Bersten2011}
{Bersten}, M.~C., {Benvenuto}, O., \& {Hamuy}, M. 2011, \apj, 729, 61

\bibitem[{{Bersten} {et~al.}(2012){Bersten}, {Benvenuto}, {Nomoto}, {Ergon},
  {Folatelli}, {Sollerman}, {Benetti}, {Botticella}, {Fraser}, {Kotak},
  {Maeda}, {Ochner}, \& {Tomasella}}]{2012Bersten}
{Bersten}, M.~C., {Benvenuto}, O.~G., {Nomoto}, K., {et~al.} 2012, \apj, 757,
  31

\bibitem[{{Bersten} {et~al.}(2018){Bersten}, {Folatelli}, {Garc{\'\i}a}, {van
  Dyk}, {Benvenuto}, {Orellana}, {Buso}, {S{\'a}nchez}, {Tanaka}, {Maeda},
  {Filippenko}, {Zheng}, {Brink}, {Cenko}, {de Jaeger}, {Kumar}, {Moriya},
  {Nomoto}, {Perley}, {Shivvers}, \& {Smith}}]{2018Bersten}
{Bersten}, M.~C., {Folatelli}, G., {Garc{\'\i}a}, F., {et~al.} 2018, \nat, 554,
  497

\bibitem[{{Bersten} {et~al.}(2013){Bersten}, {Tanaka}, {Tominaga}, {Benvenuto},
  \& {Nomoto}}]{2013Bersten}
{Bersten}, M.~C., {Tanaka}, M., {Tominaga}, N., {Benvenuto}, O.~G., \&
  {Nomoto}, K. 2013, \apj, 767, 143

\bibitem[{{Branch} \& {Wheeler}(2017)}]{2017Branch}
{Branch}, D. \& {Wheeler}, J.~C. 2017, {Supernova Explosions}

\bibitem[{{Cartier} {et~al.}(2022){Cartier}, {Hamuy}, {Contreras}, {Anderson},
  {Phillips}, {Morrell}, {Stritzinger}, {Hueichapan}, {Clocchiatti}, {Roth},
  {Thomas-Osip}, \& {Gonz{\'a}lez}}]{2022Cartier}
{Cartier}, R., {Hamuy}, M., {Contreras}, C., {et~al.} 2022, \mnras, 514, 2627

\bibitem[{{Chen} {et~al.}(2015){Chen}, {Smartt}, {Jerkstrand}, {Nicholl},
  {Bresolin}, {Kotak}, {Polshaw}, {Rest}, {Kudritzki}, {Zheng}, {Elias-Rosa},
  {Smith}, {Inserra}, {Wright}, {Kankare}, {Kangas}, \& {Fraser}}]{Chen2015}
{Chen}, T.~W., {Smartt}, S.~J., {Jerkstrand}, A., {et~al.} 2015, \mnras, 452,
  1567

\bibitem[{{Chevalier} \& {Fransson}(1994)}]{1994Chevalier}
{Chevalier}, R.~A. \& {Fransson}, C. 1994, \apj, 420, 268

\bibitem[{{Chevalier} \& {Fransson}(2008)}]{2008Chevalier}
{Chevalier}, R.~A. \& {Fransson}, C. 2008, \apjl, 683, L135

\bibitem[{{Clocchiatti} \& {Wheeler}(1997)}]{1997Clocchiatti}
{Clocchiatti}, A. \& {Wheeler}, J.~C. 1997, \apj, 491, 375

\bibitem[{{De} {et~al.}(2021){De}, {Fremling}, {Gal-Yam}, {Yaron}, {Kasliwal},
  \& {Kulkarni}}]{2021De}
{De}, K., {Fremling}, U.~C., {Gal-Yam}, A., {et~al.} 2021, \apjl, 907, L18

\bibitem[{{De} {et~al.}(2018){De}, {Kasliwal}, {Ofek}, {Moriya}, {Burke},
  {Cao}, {Cenko}, {Doran}, {Duggan}, {Fender}, {Fransson}, {Gal-Yam}, {Horesh},
  {Kulkarni}, {Laher}, {Lunnan}, {Manulis}, {Masci}, {Mazzali}, {Nugent},
  {Perley}, {Petrushevska}, {Piro}, {Rumsey}, {Sollerman}, {Sullivan}, \&
  {Taddia}}]{2018De}
{De}, K., {Kasliwal}, M.~M., {Ofek}, E.~O., {et~al.} 2018, Science, 362, 201

\bibitem[{{Dessart} {et~al.}(2018){Dessart}, {Yoon}, {Livne}, \&
  {Waldman}}]{2018Dessart}
{Dessart}, L., {Yoon}, S.-C., {Livne}, E., \& {Waldman}, R. 2018, \aap, 612,
  A61

\bibitem[{{Drout} {et~al.}(2016){Drout}, {Milisavljevic}, {Parrent},
  {Margutti}, {Kamble}, {Soderberg}, {Challis}, {Chornock}, {Fong}, {Frank},
  {Gehrels}, {Graham}, {Hsiao}, {Itagaki}, {Kasliwal}, {Kirshner}, {Macomb},
  {Marion}, {Norris}, \& {Phillips}}]{2016Drout}
{Drout}, M.~R., {Milisavljevic}, D., {Parrent}, J., {et~al.} 2016, \apj, 821,
  57

\bibitem[{{Ensman} \& {Burrows}(1992)}]{EnsmanBurrows1992}
{Ensman}, L. \& {Burrows}, A. 1992, \apj, 393, 742

\bibitem[{{Ertl} {et~al.}(2020){Ertl}, {Woosley}, {Sukhbold}, \&
  {Janka}}]{2020Ertl}
{Ertl}, T., {Woosley}, S.~E., {Sukhbold}, T., \& {Janka}, H.~T. 2020, \apj,
  890, 51

\bibitem[{{Folatelli} {et~al.}(2006){Folatelli}, {Contreras}, {Phillips},
  {Woosley}, {Blinnikov}, {Morrell}, {Suntzeff}, {Lee}, {Hamuy},
  {Gonz{\'a}lez}, {Krzeminski}, {Roth}, {Li}, {Filippenko}, {Foley},
  {Freedman}, {Madore}, {Persson}, {Murphy}, {Boissier}, {Galaz},
  {Gonz{\'a}lez}, {McCarthy}, {McWilliam}, \& {Pych}}]{Folatelli06}
{Folatelli}, G., {Contreras}, C., {Phillips}, M.~M., {et~al.} 2006, \apj, 641,
  1039

\bibitem[{{Gomez} {et~al.}(2021){Gomez}, {Berger}, {Hosseinzadeh}, {Blanchard},
  {Nicholl}, \& {Villar}}]{Gomez2021}
{Gomez}, S., {Berger}, E., {Hosseinzadeh}, G., {et~al.} 2021, \apj, 913, 143

\bibitem[{{Gonz{\'a}lez-Gait{\'a}n} {et~al.}(2015){Gonz{\'a}lez-Gait{\'a}n},
  {Tominaga}, {Molina}, {Galbany}, {Bufano}, {Anderson}, {Gutierrez},
  {F{\"o}rster}, {Pignata}, {Bersten}, {Howell}, {Sullivan}, {Carlberg}, {de
  Jaeger}, {Hamuy}, {Baklanov}, \& {Blinnikov}}]{2015Gaitan}
{Gonz{\'a}lez-Gait{\'a}n}, S., {Tominaga}, N., {Molina}, J., {et~al.} 2015,
  \mnras, 451, 2212

\bibitem[{{Guti{\'e}rrez} {et~al.}(2021){Guti{\'e}rrez}, {Bersten}, {Orellana},
  {Pastorello}, {Ertini}, {Folatelli}, {Pignata}, {Anderson}, {Smartt},
  {Sullivan}, {Pursiainen}, {Inserra}, {Elias-Rosa}, {Fraser}, {Kankare},
  {Moran}, {Reguitti}, {Reynolds}, {Stritzinger}, {Burke}, {Frohmaier},
  {Galbany}, {Hiramatsu}, {Howell}, {Kuncarayakti}, {Mattila},
  {M{\"u}ller-Bravo}, {Pellegrino}, \& {Smith}}]{Gutierrez2021}
{Guti{\'e}rrez}, C.~P., {Bersten}, M.~C., {Orellana}, M., {et~al.} 2021,
  \mnras, 504, 4907

\bibitem[{{Ho} {et~al.}(2020){Ho}, {Kulkarni}, {Perley}, {Cenko}, {Corsi},
  {Schulze}, {Lunnan}, {Sollerman}, {Gal-Yam}, {Anand}, {Barbarino}, {Bellm},
  {Bruch}, {Burns}, {De}, {Dekany}, {Delacroix}, {Duev}, {Frederiks},
  {Fremling}, {Goldstein}, {Golkhou}, {Graham}, {Hale}, {Kasliwal}, {Kupfer},
  {Laher}, {Martikainen}, {Masci}, {Neill}, {Ridnaia}, {Rusholme}, {Savchenko},
  {Shupe}, {Soumagnac}, {Strotjohann}, {Svinkin}, {Taggart}, {Tartaglia},
  {Yan}, \& {Zolkower}}]{2020Ho}
{Ho}, A. Y.~Q., {Kulkarni}, S.~R., {Perley}, D.~A., {et~al.} 2020, \apj, 902,
  86

\bibitem[{{Jacobson-Gal{\'a}n} {et~al.}(2020){Jacobson-Gal{\'a}n}, {Margutti},
  {Kilpatrick}, {Hiramatsu}, {Perets}, {Khatami}, {Foley}, {Raymond}, {Yoon},
  {Bobrick}, {Zenati}, {Galbany}, {Andrews}, {Brown}, {Cartier}, {Coppejans},
  {Dimitriadis}, {Dobson}, {Hajela}, {Howell}, {Kuncarayakti}, {Milisavljevic},
  {Rahman}, {Rojas-Bravo}, {Sand}, {Shepherd}, {Smartt}, {Stacey}, {Stroh},
  {Swift}, {Terreran}, {Vinko}, {Wang}, {Anderson}, {Baron}, {Berger},
  {Blanchard}, {Burke}, {Coulter}, {DeMarchi}, {DerKacy}, {Fremling}, {Gomez},
  {Gromadzki}, {Hosseinzadeh}, {Kasen}, {Kriskovics}, {McCully},
  {M{\"u}ller-Bravo}, {Nicholl}, {Ordasi}, {Pellegrino}, {Piro}, {P{\'a}l},
  {Ren}, {Rest}, {Rich}, {Sai}, {S{\'a}rneczky}, {Shen}, {Short}, {Siebert},
  {Stauffer}, {Szak{\'a}ts}, {Zhang}, {Zhang}, \& {Zhang}}]{2020Jacobson}
{Jacobson-Gal{\'a}n}, W.~V., {Margutti}, R., {Kilpatrick}, C.~D., {et~al.}
  2020, \apj, 898, 166

\bibitem[{{Jacobson-Gal{\'a}n} {et~al.}(2021){Jacobson-Gal{\'a}n}, {Margutti},
  {Kilpatrick}, {Raymond}, {Berger}, {Blanchard}, {Bobrick}, {Foley}, {Gomez},
  {Hosseinzadeh}, {Milisavljevic}, {Perets}, {Terreran}, \&
  {Zenati}}]{2021Jacobson}
{Jacobson-Gal{\'a}n}, W.~V., {Margutti}, R., {Kilpatrick}, C.~D., {et~al.}
  2021, \apjl, 908, L32

\bibitem[{{Jacobson-Gal{\'a}n} {et~al.}(2022){Jacobson-Gal{\'a}n},
  {Venkatraman}, {Margutti}, {Khatami}, {Terreran}, {Foley}, {Angulo}, {Angus},
  {Auchettl}, {Blanchard}, {Bobrick}, {Bright}, {Brout}, {Chambers}, {Couch},
  {Coulter}, {Clever}, {Davis}, {de Boer}, {DeMarchi}, {Dodd}, {Jones},
  {Johnson}, {Kilpatrick}, {Khetan}, {Lai}, {Langeroodi}, {Lin}, {Magnier},
  {Milisavljevic}, {Perets}, {Pierel}, {Raymond}, {Rest}, {Rest},
  {Ridden-Harper}, {Shen}, {Siebert}, {Smith}, {Taggart}, {Tinyanont},
  {Valdes}, {Villar}, {Wang}, {Yadavalli}, {Zenati}, \&
  {Zenteno}}]{2022Jacobson}
{Jacobson-Gal{\'a}n}, W.~V., {Venkatraman}, P., {Margutti}, R., {et~al.} 2022,
  \apj, 932, 58

\bibitem[{{Jerkstrand} {et~al.}(2020){Jerkstrand}, {Wongwathanarat}, {Janka},
  {Gabler}, {Alp}, {Diehl}, {Maeda}, {Larsson}, {Fransson}, {Menon}, \&
  {Heger}}]{2020Jerkstrand}
{Jerkstrand}, A., {Wongwathanarat}, A., {Janka}, H.~T., {et~al.} 2020, \mnras,
  494, 2471

\bibitem[{{Jin} {et~al.}(2021){Jin}, {Yoon}, \& {Blinnikov}}]{2021Jin}
{Jin}, H., {Yoon}, S.-C., \& {Blinnikov}, S. 2021, \apj, 910, 68

\bibitem[{{Kasen}(2010)}]{2010Kasen}
{Kasen}, D. 2010, \apj, 708, 1025

\bibitem[{{Kasen}(2017)}]{2017Kasen}
{Kasen}, D. 2017, {Unusual Supernovae and Alternative Power Sources}, ed. A.~W.
  {Alsabti} \& P.~{Murdin}, 939

\bibitem[{{Khokhlov} {et~al.}(1999){Khokhlov}, {H{\"o}flich}, {Oran},
  {Wheeler}, {Wang}, \& {Chtchelkanova}}]{1999Kok}
{Khokhlov}, A.~M., {H{\"o}flich}, P.~A., {Oran}, E.~S., {et~al.} 1999, \apjl,
  524, L107

\bibitem[{{Leloudas} {et~al.}(2012){Leloudas}, {Chatzopoulos}, {Dilday},
  {Gorosabel}, {Vinko}, {Gallazzi}, {Wheeler}, {Bassett}, {Fischer}, {Frieman},
  {Fynbo}, {Goobar}, {Jel{\'\i}nek}, {Malesani}, {Nichol}, {Nordin},
  {{\"O}stman}, {Sako}, {Schneider}, {Smith}, {Sollerman}, {Stritzinger},
  {Th{\"o}ne}, \& {de Ugarte Postigo}}]{2012Leloudas}
{Leloudas}, G., {Chatzopoulos}, E., {Dilday}, B., {et~al.} 2012, \aap, 541,
  A129

\bibitem[{{Li} {et~al.}(2020){Li}, {Wang}, {Liu}, {Wang}, {Liang}, \&
  {Dai}}]{2020Li}
{Li}, L., {Wang}, S.-Q., {Liu}, L.-D., {et~al.} 2020, \apj, 891, 98

\bibitem[{{MacFadyen} {et~al.}(2001){MacFadyen}, {Woosley}, \&
  {Heger}}]{MacFadyen2001}
{MacFadyen}, A.~I., {Woosley}, S.~E., \& {Heger}, A. 2001, \apj, 550, 410

\bibitem[{{Maeda} {et~al.}(2007){Maeda}, {Tanaka}, {Nomoto}, {Tominaga},
  {Kawabata}, {Mazzali}, {Umeda}, {Suzuki}, \& {Hattori}}]{2007Maeda}
{Maeda}, K., {Tanaka}, M., {Nomoto}, K., {et~al.} 2007, \apj, 666, 1069

\bibitem[{{Magee} \& {Maguire}(2020)}]{2020Magee}
{Magee}, M.~R. \& {Maguire}, K. 2020, \aap, 642, A189

\bibitem[{{Marcaide} {et~al.}(2009){Marcaide}, {Mart{\'\i}-Vidal}, {Alberdi},
  {P{\'e}rez-Torres}, {Ros}, {Diamond}, {Guirado}, {Lara}, {Shapiro},
  {Stockdale}, {Weiler}, {Mantovani}, {Preston}, {Schilizzi}, {Sramek},
  {Trigilio}, {van Dyk}, \& {Whitney}}]{2009Marcaide}
{Marcaide}, J.~M., {Mart{\'\i}-Vidal}, I., {Alberdi}, A., {et~al.} 2009, \aap,
  505, 927

\bibitem[{{Margalit} {et~al.}(2018){Margalit}, {Metzger}, {Thompson},
  {Nicholl}, \& {Sukhbold}}]{2018bMargalita}
{Margalit}, B., {Metzger}, B.~D., {Thompson}, T.~A., {Nicholl}, M., \&
  {Sukhbold}, T. 2018, \mnras, 475, 2659

\bibitem[{{Margutti} {et~al.}(2014){Margutti}, {Milisavljevic}, {Soderberg},
  {Chornock}, {Zauderer}, {Murase}, {Guidorzi}, {Sanders}, {Kuin}, {Fransson},
  {Levesque}, {Chandra}, {Berger}, {Bianco}, {Brown}, {Challis},
  {Chatzopoulos}, {Cheung}, {Choi}, {Chomiuk}, {Chugai}, {Contreras}, {Drout},
  {Fesen}, {Foley}, {Fong}, {Friedman}, {Gall}, {Gehrels}, {Hjorth}, {Hsiao},
  {Kirshner}, {Im}, {Leloudas}, {Lunnan}, {Marion}, {Martin}, {Morrell},
  {Neugent}, {Omodei}, {Phillips}, {Rest}, {Silverman}, {Strader},
  {Stritzinger}, {Szalai}, {Utterback}, {Vinko}, {Wheeler}, {Arnett},
  {Campana}, {Chevalier}, {Ginsburg}, {Kamble}, {Roming}, {Pritchard}, \&
  {Stringfellow}}]{2014Margutti}
{Margutti}, R., {Milisavljevic}, D., {Soderberg}, A.~M., {et~al.} 2014, \apj,
  780, 21

\bibitem[{{Mart{\'\i}-Vidal} {et~al.}(2011){Mart{\'\i}-Vidal}, {Marcaide},
  {Alberdi}, {Guirado}, {P{\'e}rez-Torres}, \& {Ros}}]{2011Marti}
{Mart{\'\i}-Vidal}, I., {Marcaide}, J.~M., {Alberdi}, A., {et~al.} 2011, \aap,
  526, A142

\bibitem[{{Martinez} {et~al.}(2020){Martinez}, {Bersten}, {Anderson},
  {Gonz{\'a}lez-Gait{\'a}n}, {F{\"o}rster}, \& {Folatelli}}]{Martinez2020}
{Martinez}, L., {Bersten}, M.~C., {Anderson}, J.~P., {et~al.} 2020, arXiv
  e-prints, arXiv:2008.05572

\bibitem[{{Mazzali} {et~al.}(2008){Mazzali}, {Valenti}, {Della Valle},
  {Chincarini}, {Sauer}, {Benetti}, {Pian}, {Piran}, {D'Elia}, {Elias-Rosa},
  {Margutti}, {Pasotti}, {Antonelli}, {Bufano}, {Campana}, {Cappellaro},
  {Covino}, {D'Avanzo}, {Fiore}, {Fugazza}, {Gilmozzi}, {Hunter}, {Maguire},
  {Maiorano}, {Marziani}, {Masetti}, {Mirabel}, {Navasardyan}, {Nomoto},
  {Palazzi}, {Pastorello}, {Panagia}, {Pellizza}, {Sari}, {Smartt},
  {Tagliaferri}, {Tanaka}, {Taubenberger}, {Tominaga}, {Trundle}, \&
  {Turatto}}]{Mazzali2008}
{Mazzali}, P.~A., {Valenti}, S., {Della Valle}, M., {et~al.} 2008, Science,
  321, 1185

\bibitem[{{Modjaz} {et~al.}(2019){Modjaz}, {Guti{\'e}rrez}, \&
  {Arcavi}}]{2019Modjaz}
{Modjaz}, M., {Guti{\'e}rrez}, C.~P., \& {Arcavi}, I. 2019, Nature Astronomy,
  3, 717

\bibitem[{{Modjaz} {et~al.}(2009){Modjaz}, {Li}, {Butler}, {Chornock},
  {Perley}, {Blondin}, {Bloom}, {Filippenko}, {Kirshner}, {Kocevski},
  {Poznanski}, {Hicken}, {Foley}, {Stringfellow}, {Berlind}, {Barrado y
  Navascues}, {Blake}, {Bouy}, {Brown}, {Challis}, {Chen}, {de Vries},
  {Dufour}, {Falco}, {Friedman}, {Ganeshalingam}, {Garnavich}, {Holden},
  {Illingworth}, {Lee}, {Liebert}, {Marion}, {Olivier}, {Prochaska},
  {Silverman}, {Smith}, {Starr}, {Steele}, {Stockton}, {Williams}, \&
  {Wood-Vasey}}]{2009Modjaz}
{Modjaz}, M., {Li}, W., {Butler}, N., {et~al.} 2009, \apj, 702, 226

\bibitem[{{Moriya} {et~al.}(2013){Moriya}, {Blinnikov}, {Baklanov}, {Sorokina},
  \& {Dolgov}}]{2013Moriya}
{Moriya}, T.~J., {Blinnikov}, S.~I., {Baklanov}, P.~V., {Sorokina}, E.~I., \&
  {Dolgov}, A.~D. 2013, \mnras, 430, 1402

\bibitem[{{Nakar}(2015)}]{2015Nakar}
{Nakar}, E. 2015, \apj, 807, 172

\bibitem[{{Nakar} \& {Piro}(2014)}]{Nakar2014}
{Nakar}, E. \& {Piro}, A.~L. 2014, \apj, 788, 193

\bibitem[{{Nicholl} {et~al.}(2017){Nicholl}, {Guillochon}, \&
  {Berger}}]{2017Nicholl}
{Nicholl}, M., {Guillochon}, J., \& {Berger}, E. 2017, \apj, 850, 55

\bibitem[{{Nicholl} \& {Smartt}(2016)}]{2016Nicholl}
{Nicholl}, M. \& {Smartt}, S.~J. 2016, \mnras, 457, L79

\bibitem[{{Nicholl} {et~al.}(2015){Nicholl}, {Smartt}, {Jerkstrand}, {Sim},
  {Inserra}, {Anderson}, {Baltay}, {Benetti}, {Chambers}, {Chen}, {Elias-Rosa},
  {Feindt}, {Flewelling}, {Fraser}, {Gal-Yam}, {Galbany}, {Huber}, {Kangas},
  {Kankare}, {Kotak}, {Kr{\"u}hler}, {Maguire}, {McKinnon}, {Rabinowitz},
  {Rostami}, {Schulze}, {Smith}, {Sullivan}, {Tonry}, {Valenti}, \&
  {Young}}]{2015Nicholla}
{Nicholl}, M., {Smartt}, S.~J., {Jerkstrand}, A., {et~al.} 2015, \apjl, 807,
  L18

\bibitem[{{Nishimura} {et~al.}(2015){Nishimura}, {Takiwaki}, \&
  {Thielemann}}]{2015Nishimura}
{Nishimura}, N., {Takiwaki}, T., \& {Thielemann}, F.-K. 2015, \apj, 810, 109

\bibitem[{{Noebauer} {et~al.}(2017){Noebauer}, {Kromer}, {Taubenberger},
  {Baklanov}, {Blinnikov}, {Sorokina}, \& {Hillebrandt}}]{2017Noebauer}
{Noebauer}, U.~M., {Kromer}, M., {Taubenberger}, S., {et~al.} 2017, \mnras,
  472, 2787

\bibitem[{{Nomoto} \& {Hashimoto}(1988)}]{1988Ken}
{Nomoto}, K. \& {Hashimoto}, M. 1988, Physics Reports, 163, 13

\bibitem[{{Orellana} \& {Bersten}(2021)}]{2021BAAA}
{Orellana}, M. \& {Bersten}, M.~C. 2021, Boletin de la Asociacion Argentina de
  Astronomia La Plata Argentina, 62, 89

\bibitem[{{Ouyed} {et~al.}(2016){Ouyed}, {Leahy}, \& {Koning}}]{2016quark}
{Ouyed}, R., {Leahy}, D., \& {Koning}, N. 2016, \apj, 818, 77

\bibitem[{{Paxton} {et~al.}(2011){Paxton}, {Bildsten}, {Dotter}, {Herwig},
  {Lesaffre}, \& {Timmes}}]{2011Paxton}
{Paxton}, B., {Bildsten}, L., {Dotter}, A., {et~al.} 2011, \apjs, 192, 3

\bibitem[{{Piro}(2015)}]{2015Piro}
{Piro}, A.~L. 2015, \apjl, 808, L51

\bibitem[{{Rabinak} \& {Waxman}(2011)}]{2011Rabinak}
{Rabinak}, I. \& {Waxman}, E. 2011, \apj, 728, 63

\bibitem[{{Ray} {et~al.}(1993){Ray}, {Singh}, \& {Sutaria}}]{1993Ray}
{Ray}, A., {Singh}, K.~P., \& {Sutaria}, F.~K. 1993, Journal of Astrophysics
  and Astronomy, 14, 53

\bibitem[{{Senno} {et~al.}(2016){Senno}, {Murase}, \&
  {M{\'e}sz{\'a}ros}}]{2016Senno}
{Senno}, N., {Murase}, K., \& {M{\'e}sz{\'a}ros}, P. 2016, \prd, 93, 083003

\bibitem[{{Shankar} {et~al.}(2021){Shankar}, {M{\"o}sta}, {Barnes}, {Duffell},
  \& {Kasen}}]{2021Shankar}
{Shankar}, S., {M{\"o}sta}, P., {Barnes}, J., {Duffell}, P.~C., \& {Kasen}, D.
  2021, \mnras, 508, 5390

\bibitem[{{Sharon} \& {Kushnir}(2020)}]{2020Sharon}
{Sharon}, A. \& {Kushnir}, D. 2020, \mnras, 496, 4517

\bibitem[{{Shen} {et~al.}(2019){Shen}, {Quataert}, \& {Pakmor}}]{2019Shen}
{Shen}, K.~J., {Quataert}, E., \& {Pakmor}, R. 2019, \apj, 887, 180

\bibitem[{{Smith} {et~al.}(2016){Smith}, {Sullivan}, {D'Andrea}, {Castander},
  {Casas}, {Prajs}, {Papadopoulos}, {Nichol}, {Karpenka}, {Bernard}, {Brown},
  {Cartier}, {Cooke}, {Curtin}, {Davis}, {Finley}, {Foley}, {Gal-Yam},
  {Goldstein}, {Gonz{\'a}lez-Gait{\'a}n}, {Gupta}, {Howell}, {Inserra},
  {Kessler}, {Lidman}, {Marriner}, {Nugent}, {Pritchard}, {Sako}, {Smartt},
  {Smith}, {Spinka}, {Thomas}, {Wolf}, {Zenteno}, {Abbott}, {Benoit-L{\'e}vy},
  {Bertin}, {Brooks}, {Buckley-Geer}, {Carnero Rosell}, {Carrasco Kind},
  {Carretero}, {Crocce}, {Cunha}, {da Costa}, {Desai}, {Diehl}, {Doel},
  {Estrada}, {Evrard}, {Flaugher}, {Fosalba}, {Frieman}, {Gerdes}, {Gruen},
  {Gruendl}, {James}, {Kuehn}, {Kuropatkin}, {Lahav}, {Li}, {Marshall},
  {Martini}, {Miller}, {Miquel}, {Nord}, {Ogando}, {Plazas}, {Reil}, {Romer},
  {Roodman}, {Rykoff}, {Sanchez}, {Scarpine}, {Schubnell}, {Sevilla-Noarbe},
  {Soares-Santos}, {Sobreira}, {Suchyta}, {Swanson}, {Tarle}, {Walker},
  {Wester}, \& {DES Collaboration}}]{2016Smith}
{Smith}, M., {Sullivan}, M., {D'Andrea}, C.~B., {et~al.} 2016, \apjl, 818, L8

\bibitem[{{Sobacchi} {et~al.}(2017){Sobacchi}, {Granot}, {Bromberg}, \&
  {Sormani}}]{2017Sobacchi}
{Sobacchi}, E., {Granot}, J., {Bromberg}, O., \& {Sormani}, M.~C. 2017, \mnras,
  472, 616

\bibitem[{{Soderberg} {et~al.}(2008){Soderberg}, {Berger}, {Page}, {Schady},
  {Parrent}, {Pooley}, {Wang}, {Ofek}, {Cucchiara}, {Rau}, {Waxman}, {Simon},
  {Bock}, {Milne}, {Page}, {Barentine}, {Barthelmy}, {Beardmore}, {Bietenholz},
  {Brown}, {Burrows}, {Burrows}, {Byrngelson}, {Cenko}, {Chandra}, {Cummings},
  {Fox}, {Gal-Yam}, {Gehrels}, {Immler}, {Kasliwal}, {Kong}, {Krimm},
  {Kulkarni}, {Maccarone}, {M{\'e}sz{\'a}ros}, {Nakar}, {O'Brien}, {Overzier},
  {de Pasquale}, {Racusin}, {Rea}, \& {York}}]{2008Soderberg}
{Soderberg}, A.~M., {Berger}, E., {Page}, K.~L., {et~al.} 2008, \nat, 453, 469

\bibitem[{{Soker} \& {Gilkis}(2017)}]{2017Soker}
{Soker}, N. \& {Gilkis}, A. 2017, \apj, 851, 95

\bibitem[{{Stritzinger} {et~al.}(2002){Stritzinger}, {Hamuy}, {Suntzeff},
  {Smith}, {Phillips}, {Maza}, {Strolger}, {Antezana}, {Gonz{\'a}lez},
  {Wischnjewsky}, {Candia}, {Espinoza}, {Gonz{\'a}lez}, {Stubbs}, {Becker},
  {Rubenstein}, \& {Galaz}}]{2002Stritzinger}
{Stritzinger}, M., {Hamuy}, M., {Suntzeff}, N.~B., {et~al.} 2002, \aj, 124,
  2100

\bibitem[{{Stritzinger} {et~al.}(2020){Stritzinger}, {Taddia}, {Holmbo},
  {Baron}, {Contreras}, {Karamehmetoglu}, {Phillips}, {Sollerman}, {Suntzeff},
  {Vinko}, {Ashall}, {Avila}, {Burns}, {Campillay}, {Castellon}, {Folatelli},
  {Galbany}, {Hoeflich}, {Hsiao}, {Marion}, {Morrell}, \&
  {Wheeler}}]{2020Stritzinger}
{Stritzinger}, M.~D., {Taddia}, F., {Holmbo}, S., {et~al.} 2020, \aap, 634, A21

\bibitem[{{Swartz} {et~al.}(1995){Swartz}, {Sutherland}, \&
  {Harkness}}]{1995Sutherland}
{Swartz}, D.~A., {Sutherland}, P.~G., \& {Harkness}, R.~P. 1995, \apj, 446, 766

\bibitem[{{Taddia} {et~al.}(2016){Taddia}, {Fremling}, {Sollerman}, {Corsi},
  {Gal-Yam}, {Karamehmetoglu}, {Lunnan}, {Bue}, {Ergon}, {Kasliwal},
  {Vreeswijk}, \& {Wozniak}}]{2016Taddia}
{Taddia}, F., {Fremling}, C., {Sollerman}, J., {et~al.} 2016, \aap, 592, A89

\bibitem[{{Taddia} {et~al.}(2018){Taddia}, {Sollerman}, {Fremling},
  {Karamehmetoglu}, {Quimby}, {Gal-Yam}, {Yaron}, {Kasliwal}, {Kulkarni},
  {Nugent}, {Smadja}, \& {Tao}}]{Taddia18}
{Taddia}, F., {Sollerman}, J., {Fremling}, C., {et~al.} 2018, \aap, 609, A106

\bibitem[{{Tanaka} {et~al.}(2009{\natexlab{a}}){Tanaka}, {Kawabata}, {Maeda},
  {Iye}, {Hattori}, {Pian}, {Nomoto}, {Mazzali}, \&
  {Tominaga}}]{2009Tanaka_05bf}
{Tanaka}, M., {Kawabata}, K.~S., {Maeda}, K., {et~al.} 2009{\natexlab{a}},
  \apj, 699, 1119

\bibitem[{{Tanaka} {et~al.}(2009{\natexlab{b}}){Tanaka}, {Tominaga}, {Nomoto},
  {Valenti}, {Sahu}, {Minezaki}, {Yoshii}, {Yoshida}, {Anupama}, {Benetti},
  {Chincarini}, {Della Valle}, {Mazzali}, \& {Pian}}]{2009Tanaka_08D}
{Tanaka}, M., {Tominaga}, N., {Nomoto}, K., {et~al.} 2009{\natexlab{b}}, \apj,
  692, 1131

\bibitem[{{Tanaka} {et~al.}(2009{\natexlab{c}}){Tanaka}, {Yamanaka}, {Maeda},
  {Kawabata}, {Hattori}, {Minezaki}, {Valenti}, {Della Valle}, {Sahu},
  {Anupama}, {Tominaga}, {Nomoto}, {Mazzali}, \& {Pian}}]{2009Tanaka_b}
{Tanaka}, M., {Yamanaka}, M., {Maeda}, K., {et~al.} 2009{\natexlab{c}}, \apj,
  700, 1680

\bibitem[{{Tominaga} {et~al.}(2005){Tominaga}, {Tanaka}, {Nomoto}, {Mazzali},
  {Deng}, {Maeda}, {Umeda}, {Modjaz}, {Hicken}, {Challis}, {Kirshner},
  {Wood-Vasey}, {Blake}, {Bloom}, {Skrutskie}, {Szentgyorgyi}, {Falco},
  {Inada}, {Minezaki}, {Yoshii}, {Kawabata}, {Iye}, {Anupama}, {Sahu}, \&
  {Prabhu}}]{2005Tominaga}
{Tominaga}, N., {Tanaka}, M., {Nomoto}, K., {et~al.} 2005, \apjl, 633, L97

\bibitem[{{Vreeswijk} {et~al.}(2017){Vreeswijk}, {Leloudas}, {Gal-Yam}, {De
  Cia}, {Perley}, {Quimby}, {Waldman}, {Sullivan}, {Yan}, {Ofek}, {Fremling},
  {Taddia}, {Sollerman}, {Valenti}, {Arcavi}, {Howell}, {Filippenko}, {Cenko},
  {Yaron}, {Kasliwal}, {Cao}, {Ben-Ami}, {Horesh}, {Rubin}, {Lunnan}, {Nugent},
  {Laher}, {Rebbapragada}, {Wo{\'z}niak}, \& {Kulkarni}}]{2017Vreeswijk}
{Vreeswijk}, P.~M., {Leloudas}, G., {Gal-Yam}, A., {et~al.} 2017, \apj, 835, 58

\bibitem[{{Vurm} \& {Metzger}(2021)}]{2021magnetar}
{Vurm}, I. \& {Metzger}, B.~D. 2021, \apj, 917, 77

\bibitem[{{Wang} {et~al.}(2015){Wang}, {Wang}, {Dai}, \& {Wu}}]{2015Wang}
{Wang}, S.~Q., {Wang}, L.~J., {Dai}, Z.~G., \& {Wu}, X.~F. 2015, \apj, 799, 107

\bibitem[{{Wheeler} {et~al.}(2015){Wheeler}, {Johnson}, \&
  {Clocchiatti}}]{2015Wheeler}
{Wheeler}, J.~C., {Johnson}, V., \& {Clocchiatti}, A. 2015, \mnras, 450, 1295

\bibitem[{{Wilk} {et~al.}(2019){Wilk}, {Hillier}, \& {Dessart}}]{2019Wilk}
{Wilk}, K.~D., {Hillier}, D.~J., \& {Dessart}, L. 2019, \mnras, 487, 1218

\bibitem[{{Woosley} \& {Bloom}(2006)}]{2006Woosley}
{Woosley}, S.~E. \& {Bloom}, J.~S. 2006, \araa, 44, 507

\bibitem[{{Woosley} {et~al.}(1994){Woosley}, {Eastman}, {Weaver}, \&
  {Pinto}}]{Woosley1994}
{Woosley}, S.~E., {Eastman}, R.~G., {Weaver}, T.~A., \& {Pinto}, P.~A. 1994,
  \apj, 429, 300

\bibitem[{{Woosley} {et~al.}(2021){Woosley}, {Sukhbold}, \&
  {Kasen}}]{2021Woosley}
{Woosley}, S.~E., {Sukhbold}, T., \& {Kasen}, D.~N. 2021, \apj, 913, 145

\bibitem[{{Yang} {et~al.}(2021){Yang}, {Sollerman}, {Chen}, {Kool}, {Lunnan},
  {Schulze}, {Strotjohann}, {Horesh}, {Kasliwal}, {Kupfer}, {Mahabal}, {Masci},
  {Nugent}, {Perley}, {Riddle}, {Rusholme}, \& {Sharma}}]{2021Yang}
{Yang}, S., {Sollerman}, J., {Chen}, T.~W., {et~al.} 2021, \aap, 646, A22

\bibitem[{{Yao} {et~al.}(2020){Yao}, {De}, {Kasliwal}, {Ho}, {Schulze}, {Li},
  {Kulkarni}, {Fruchter}, {Rubin}, {Perley}, {Fuller}, {Piro}, {Fremling},
  {Bellm}, {Burruss}, {Duev}, {Feeney}, {Gal-Yam}, {Golkhou}, {Graham},
  {Helou}, {Kupfer}, {Laher}, {Masci}, {Miller}, {Rusholme}, {Shupe}, {Smith},
  {Sollerman}, {Soumagnac}, \& {Zolkower}}]{2020Yao}
{Yao}, Y., {De}, K., {Kasliwal}, M.~M., {et~al.} 2020, \apj, 900, 46

\bibitem[{{Zs{\'\i}ros} {et~al.}(2022){Zs{\'\i}ros}, {Nagy}, \&
  {Szalai}}]{2022Zsiros}
{Zs{\'\i}ros}, S., {Nagy}, A.~P., \& {Szalai}, T. 2022, \mnras, 509, 3235

\end{thebibliography}

\end{document}